\documentclass[sigconf,authorversion]{acmart}

\setcopyright{rightsretained}
\copyrightyear{2025}
\acmYear{2025}
\acmConference{SIGGRAPH Talks '25}{August 10-14, 2025}{Vancouver, BC, Canada}\acmBooktitle{Special Interest Group on Computer Graphics and Interactive Techniques Conference Talks (SIGGRAPH Talks '25), August 10-14, 2025}\acmDOI{10.1145/3721239.3734093}
\acmISBN{979-8-4007-1541-9/2025/08}

\usepackage{tikz}
\usepackage{algorithm}
\usepackage{algpseudocode}
\usepackage{pgfplots}
\usepackage{contour}
\usepgfplotslibrary{patchplots}
\usetikzlibrary{shadows,backgrounds,arrows,positioning,matrix,calc,decorations.text}
\pgfplotsset{compat=newest}

\definecolor{myred}{rgb}{0.86,0.00,0.00}
\definecolor{myredlight}{rgb}{0.97,0.75,0.75}
\definecolor{myredlighter}{rgb}{0.99,0.94,0.94}
\definecolor{myredlighterr}{rgb}{1.0,0.98,0.98}
\definecolor{myblue}{rgb}{0.00,0.20,0.70}
\definecolor{mybluelight}{rgb}{0.75,0.80,0.93}
\definecolor{mybluelighter}{rgb}{0.94,0.95,0.98}
\definecolor{mybluelighterr}{rgb}{0.98,0.99,1.0}
\definecolor{mygreen}{rgb}{0.10,0.50,0.10}
\definecolor{mygreenlight}{rgb}{0.78,0.88,0.78}
\definecolor{mygreenlighter}{rgb}{0.94,0.97,0.94}
\definecolor{mygreenlighterr}{rgb}{0.99,0.99,0.99}
\definecolor{mygrey}{rgb}{0.40,0.40,0.40}
\definecolor{mygreylight}{rgb}{0.85,0.85,0.85}
\definecolor{mygreylighter}{rgb}{0.96,0.96,0.96}
\definecolor{mygreylighterr}{rgb}{0.99,0.99,0.99}
\definecolor{myorange}{rgb}{1.0,0.50,0.00}
\definecolor{myorangelight}{rgb}{1.0,0.87,0.75}
\definecolor{myorangelighter}{rgb}{1.0,0.96,0.93}
\definecolor{myorangelighterr}{rgb}{1.0,0.99,0.98}

\begin{document}

\title{Adaptive Tetrahedral Grids for Volumetric Path-Tracing}

\author{Anis Benyoub}
\orcid{0009-0007-4609-1728}
\affiliation{%
  \institution{Intel Corporation}
  \city{Paris}
  \country{France}
}

\author{Jonathan Dupuy}
\orcid{0000-0002-4447-3147}
\affiliation{%
  \institution{Intel Corporation}
  \city{Grenoble}
  \country{France}
}

\begin{abstract}
    We advertise the use of tetrahedral grids constructed via the 
    longest edge bisection algorithm for rendering volumetric data with 
    path tracing. The key benefits of such grids is two-fold. First, they 
    provide a highly adaptive space-partitioning representation that limits 
    the memory footprint of volumetric assets. Second, each (tetrahedral) 
    cell has exactly 4 neighbors within the volume (one per face of 
    each tetrahedron) or less at boundaries. We leverage these properties 
    to devise optimized algorithms and data-structures to compute and path-trace 
    adaptive tetrahedral grids on the GPU. In practice, our GPU implementation 
    outperforms regular grids by up to $\times$30 and renders 
    production assets in realtime at 32 samples per pixel.
\end{abstract}

\begin{teaserfigure}
  \begin{tikzpicture}
    \setlength{\fboxsep}{0pt}
    \setlength{\fboxrule}{0.7pt}
    \begin{scope}[shift={(12.0, 0.0)}]
        \begin{scope}
            \node[black, draw=none, anchor = center] at (3.0, -0.30) {\scalebox{1.0}{regular grid}};
            \node[black, draw=none, anchor = center] at (3.0, -0.75) {\scalebox{1.0}{1024 $\times$ 1024, 1024spp}};
            \node[black, draw=none, anchor = center] at (3.0, -1.15) {\scalebox{1.0}{render time: \textbf{\color{myred}49.773s}}};
            \node[anchor=south west] at (0, 0) {
              \fbox{\includegraphics[trim = 0 100 100 100, clip, width=0.32\textwidth]{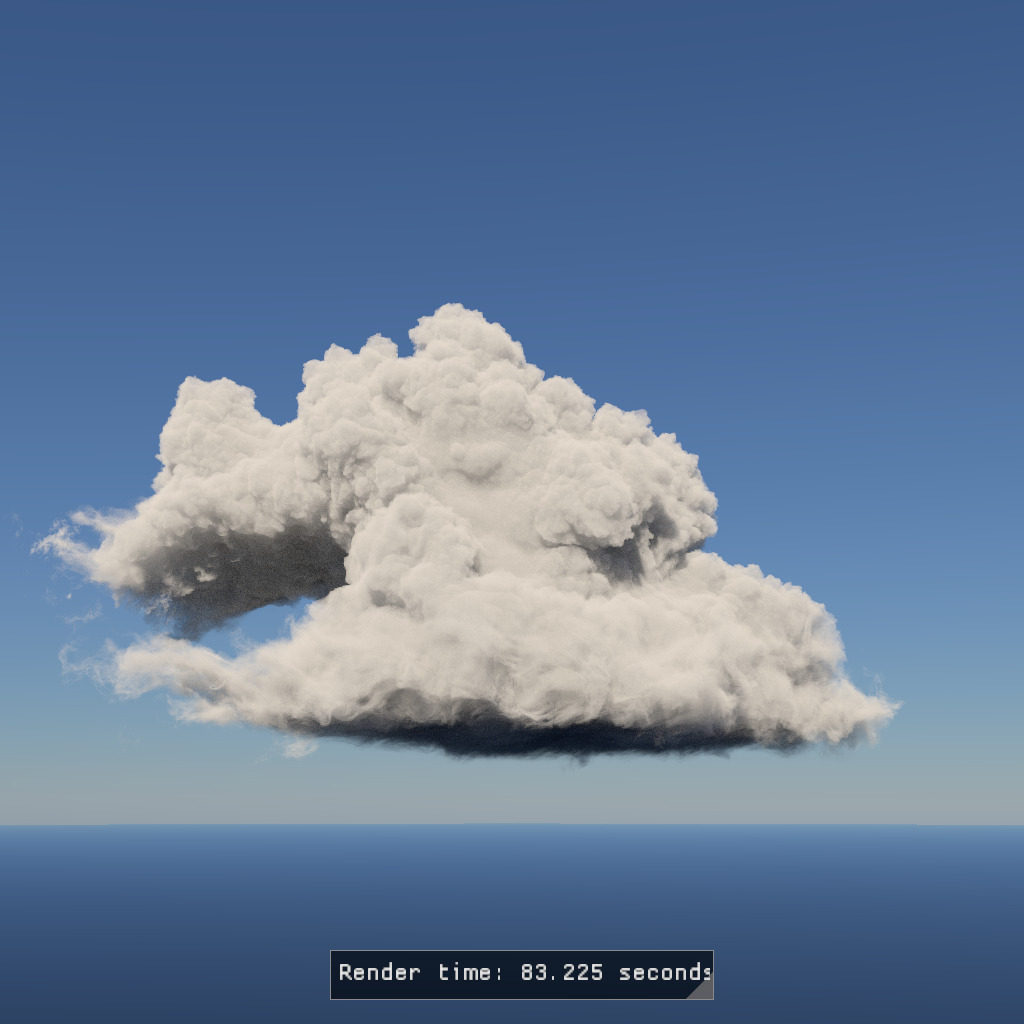}}
              };
            \node[white, draw=none, anchor = center] at (3.0, 4.9) {\scalebox{1.0}{(reference)}};
        \end{scope}
    \end{scope}
    \begin{scope}[shift={(6.0, 0.0)}]
        \begin{scope}
            \node[black, draw=none, anchor = center] at (3.0, -0.30) {\scalebox{1.0}{tetrahedral grid}};
            \node[black, draw=none, anchor = center] at (3.0, -0.75) {\scalebox{1.0}{1024 $\times$ 1024, 1024spp}};
            \node[black, draw=none, anchor = center] at (3.0, -1.15) {\scalebox{1.0}{render time: \textbf{\color{mygreen}1.633s}}};
            \node[anchor=south west] at (0, 0) {
                \fbox{\includegraphics[trim = 0 100 100 100, clip, width=0.32\textwidth]{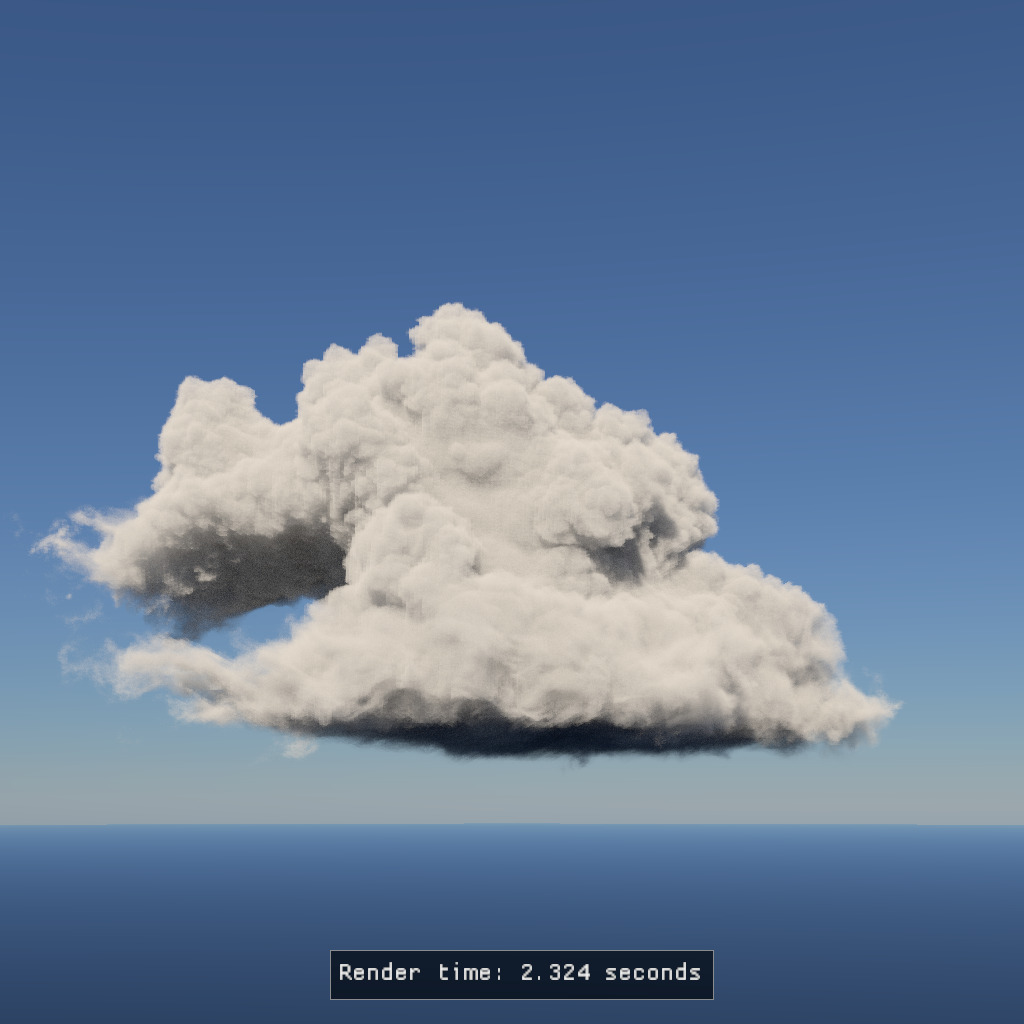}}
            };
            \node[white, draw=none, anchor = center] at (3.0, 4.9) {\scalebox{1.0}{(ours)}};
        \end{scope}
    \end{scope}
    \begin{scope}[shift={(0.0, 0.0)}]
        \begin{scope}
            \node[black, draw=none, anchor = center] at (3.0, -0.30) {\scalebox{1.0}{tetrahedral grid}};
            \node[anchor=south west] at (0, 0) {
              \includegraphics[trim = 20 100 10 215, clip, width=0.32\textwidth]{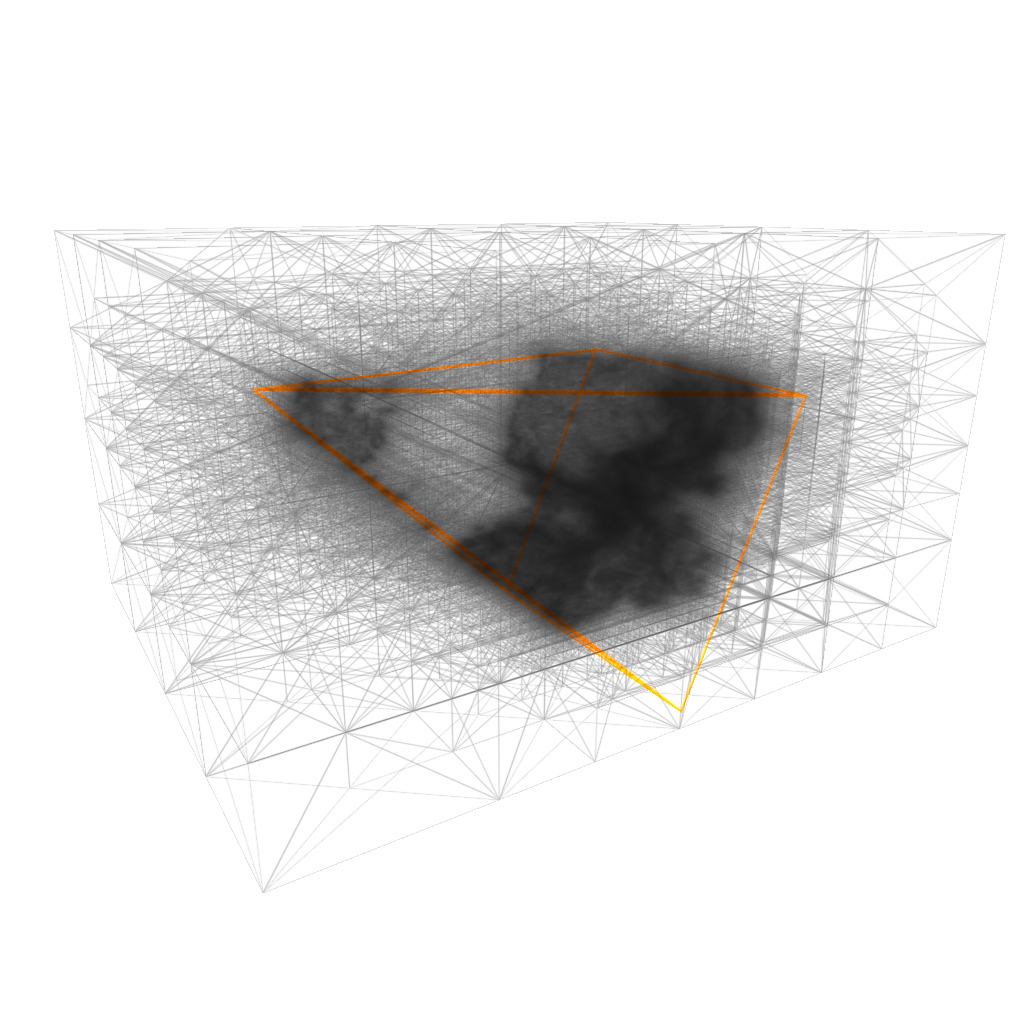}
            };
        \end{scope}
    \end{scope}
\end{tikzpicture}
  \caption{We advertise the use of adaptive tetrahedral grids for efficiently path-tracing volumetric data. Our rendering 
  pipeline starts by converting OpenVDB data into (left) an adaptive tetrahedral grid that targets pixel level 
  voxels within the camera frustum. We then path-trace this grid on the GPU to produce (center) our final 
  render, which is visually similar to (right) a regular grid while rendering up to $\times$30 times faster. 
  }
  \label{fig_teaser}
\end{teaserfigure}
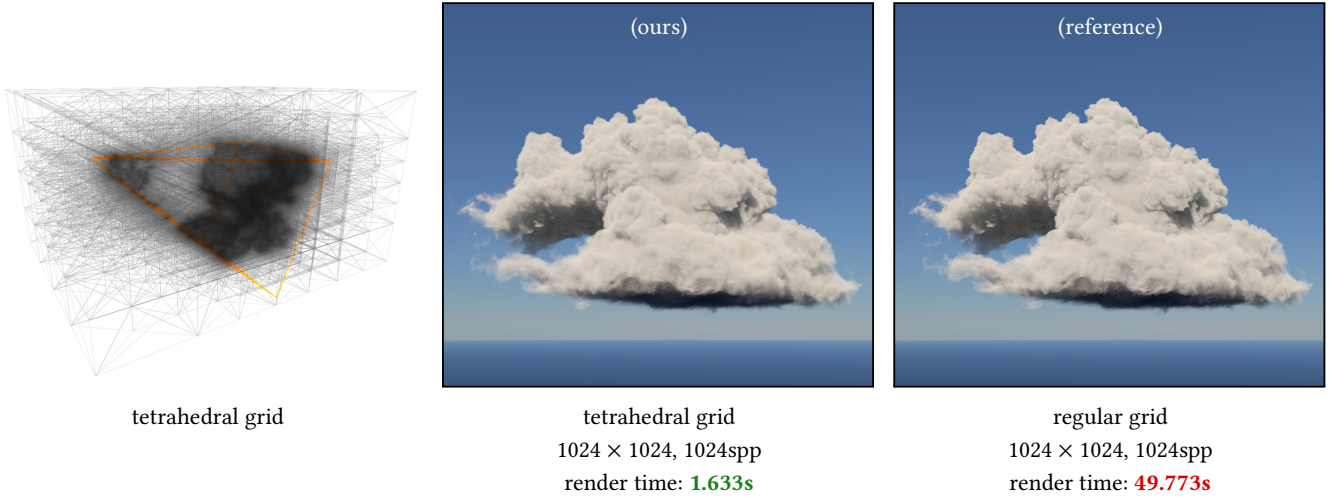

\maketitle

\section{Introduction}
\label{sec_introduction}
A tetrahedral grid partitions space with tetrahedrons rather than cubes (or cuboids). 
An important advantage of using tetrahedrons over cubes is that it is possible to build 
grids such as the one shown in Figure~\ref{fig_teaser} on the left, which have the unique property 
of being both:
\begin{enumerate}
\item \emph{adaptive}, i.e., the volume of each cell can vary throughout the grid,
\item \emph{conforming}, i.e., each cell has exactly one tetrahedral neighbor 
per face (or zero if the face lies at a boundary). Thus, each cell has exactly 
4 neighbors (or less if it lies at a boundary) independently of its volume.
\end{enumerate}
This key property has made such grids a ubiquitous numerical computation tool for, e.g., 
partial differential equations~\cite{Rivara1991}, fluid simulation~\cite{Ando2013}, and 
isosurface extraction~\cite{Pascucci2004,Weiss2008,Scholz2015}. 

In this talk, we wish to show that such grids are also very relevant for path 
tracing volumetric data. Indeed, their regularity makes them ideal for computing 
ray-marching operations on the GPU, which allows to path-trace production assets 
as that shown in Figure~\ref{fig_teaser} in realtime. We share the algorithms that we 
tailored for generating and path-tracing adaptive tetrahedral grids and showcase their 
efficiency with renderings and performance measurements.

\section{Grid Construction}
\label{sec_leb}
\paragraph{Longest Edge Bisection Algorithm}
We build our adaptive tetrahedral grids via our own implementation of the longest 
edge bisection (LEB) algorithm described by Maubach~\shortcite{Maubach1995}. 
We start from a control cube that we tessellate into 24 tetrahedrons (one per halfedge 
of the cube) with neighboring information. Each tetrahedron is composed of the two vertices 
of its associated halfedge, a face point located at the center of the face carrying the halfedge, 
and a volume point located at the center of the cube. 
We recursively bisect these initial tetrahedrons until our subdivision criteria is met 
(we describe our criteria in the next section). The bisection of any tetrahedron propagates 
to neighboring ones according to a deterministic rule that guarantees a conforming 
tetrahedral grid.

\paragraph{Subdivision Criteria}
We build our grids by taking OpenVDB~\cite{OpenVDB2013} data as input. We place our control cube 
so as to match the boundaries of the OpenVDB data and start recursively bisecting tetrahedrons based 
on two main criteria:
\begin{enumerate}
  \item \emph{Density-Variation.} We only bisect a tetrahedron whenever the density present within its 
  volume varies above a certain threshold. In practice, we use \mbox{$\frac{\max(D) - \min(D)}{\textrm{mean}(D)}$} 
  where $D$ denotes all density values present in a cell's volume. 
  This approach allows to increase the resolution of our grid only at locations where 
  more detail is present in the volume.
  \item \emph{Camera-dependancy.} We also rely on camera parameters to prevent subdivision for 
  tetrahedrons that either lie outside of the view frustum, or project into less than one pixel 
  in the final render. 
\end{enumerate}
We show the result of our subdividion criteria in Figure~\ref{fig_teaser}. 

\section{Volumetric Path Tracing}
\label{sec_volpath}

\paragraph{Rendering Algorithm}
We render our volumes using a Monte Carlo path tracer. Our path tracer solves light transport within 
heterogeneous media as shown in Algorithm~\ref{alg_trace}. The media is represented via a volumetric density 
value per tetrahedron within our grid. Note that our implementation also supports extra channels 
for, e.g., temperature, albedo, phase function parameters, etc.

\begin{algorithm}[H]
  \footnotesize
  \begin{algorithmic}[1]
      \caption{Volumetric Path-Tracing Algorithm}
      \label{alg_trace}
      \Function{Trace}{ray: Ray}
        \If {\Call{IntersectGrid}{ray}}
          \State ray.origin $\gets$ \Call{GridIntersectionPoint}{ray}
          \While {InsideGrid(ray)}
            \State $\lambda$ $\gets$ \Call{ReadDensity}{ray.origin}
            \State $\ell$ $\gets$ \Call{SampleFreePath}{$\lambda$}
            \State ray.origin $\gets$ \Call{MarchThroughCells}{ray, $\ell$}
            \State ray.direction $\gets$ \Call{SamplePhaseFunction}{ray.direction}
          \EndWhile
        \EndIf
        \State \Return ray
      \EndFunction
  \end{algorithmic}
\end{algorithm}

\paragraph{Path Tracing within Tetrahedral Grids}
The main component of our renderer is a tracing algorithm that allows to simulate free paths within our 
tetrahedral grids. Since each cell has at most four neighbors, we walk through the grid very efficiently 
by evaluating at most 4 ray/tetrahedron intersections per cell. To further speed-up this computation, 
we store the normals of the faces of each tetrahedron. Thanks to this information, we only need to test at 
most 3 faces by using the direction of the ray. In practice, only a few bits are enough to store these 
normals as our tetrahedral grids only produce 18 different face orientations.

\section{Results}
\label{sec_results}
\begin{figure}
  \begin{tikzpicture}
    \setlength{\fboxsep}{0pt}
    \setlength{\fboxrule}{0.7pt}
    \begin{scope}[shift={(0.0, 0.0)}]
        \begin{scope}
          \node[black, draw=none, anchor = center] at (0.0, -2.20) {\scalebox{1.0}{tetrahedral grid}};
          \node[black, draw=none, anchor = center] at (0.0, -2.6) {\scalebox{1.0}{1024$\times$ 1024, 16kspp}};
          \node[black, draw=none, anchor = center] at (0.0, -3.0) {\scalebox{1.0}{render time: \textbf{\color{mygreen}69.8s}}};
          \node[anchor=center] at (0, 0) {
            \fbox{\includegraphics[trim = 0 200 100 0, clip, width=0.23\textwidth]{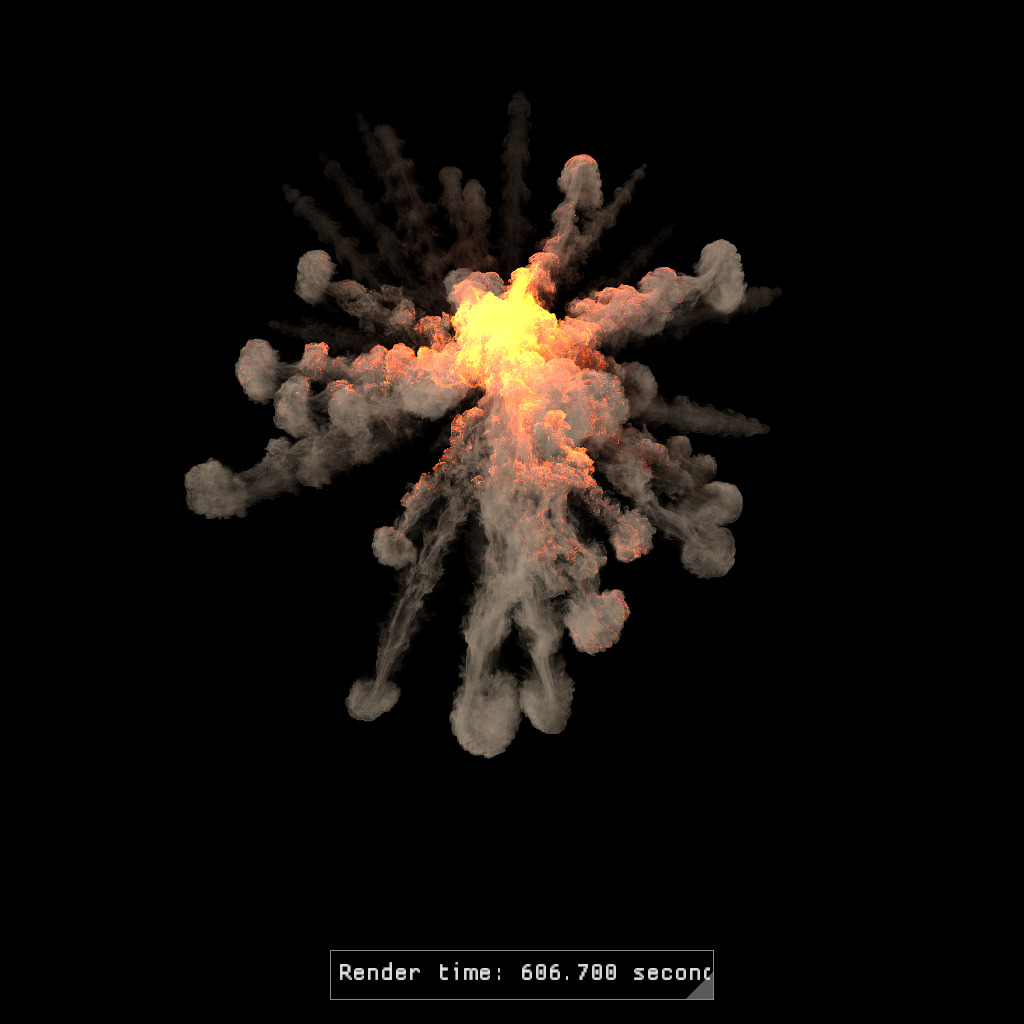}}
            };
          \node[white, draw=none, anchor = center] at (0.0, 1.65) {\scalebox{1.0}{(ours)}};  
          \end{scope}
    \end{scope}
    \begin{scope}[shift={(4.2, 0.0)}]
      \begin{scope}
          \node[black, draw=none, anchor = center] at (0.0, -2.2) {\scalebox{1.0}{regular grid}};
          \node[black, draw=none, anchor = center] at (0.0, -2.6) {\scalebox{1.0}{1024$\times$ 1024, 16kspp}};
          \node[black, draw=none, anchor = center] at (0.0, -3.0) {\scalebox{1.0}{render time: \textbf{\color{myred}606.7s}}};
          \node[anchor=center] at (0, 0) {
              \fbox{\includegraphics[trim = 0 200 100 0, clip, width=0.23\textwidth]{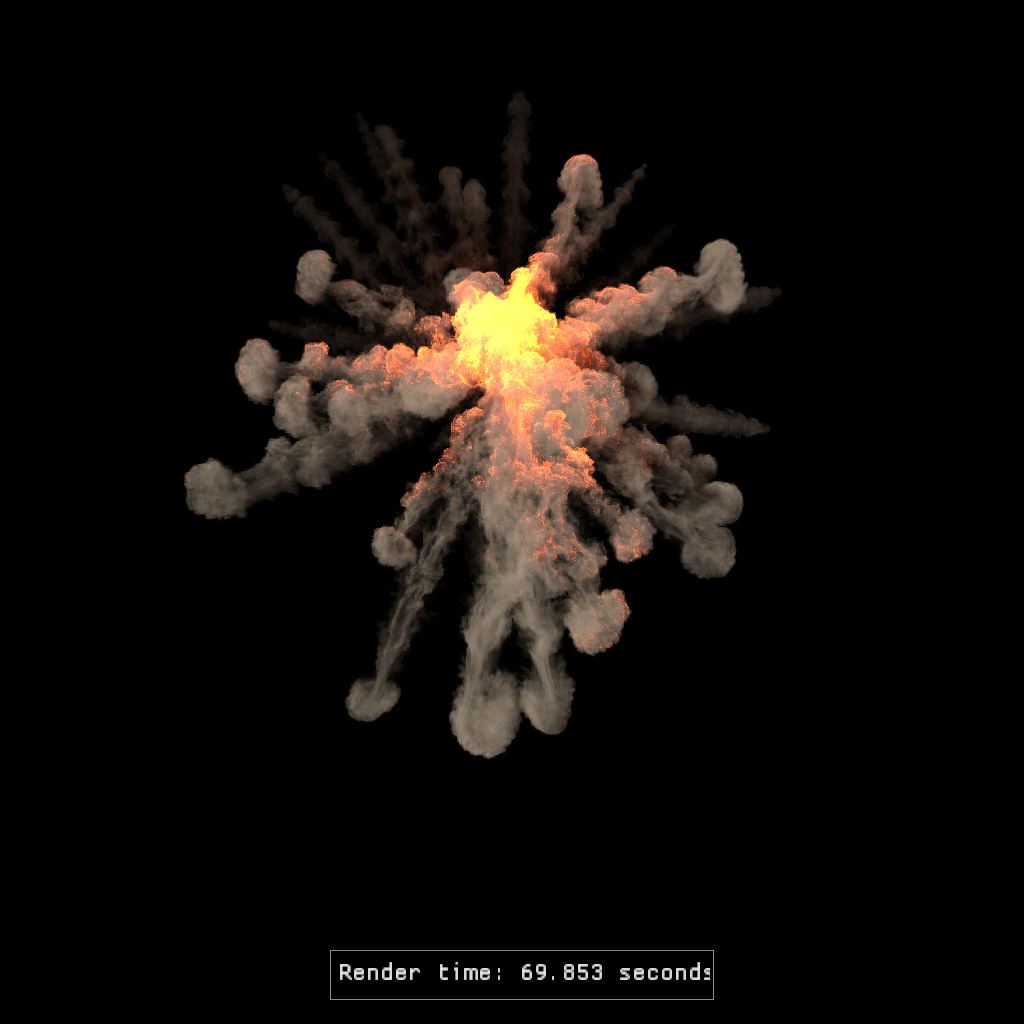}}
          };
          \node[white, draw=none, anchor = center] at (0.0, 1.65) {\scalebox{1.0}{(reference)}};  
      \end{scope}
    \end{scope}  
  \end{tikzpicture}
  \caption{Rendering of a volumetric asset with density and temperature parameters. See our accompanying video for the animated sequence.}
  \label{fig_explosion}
\end{figure}
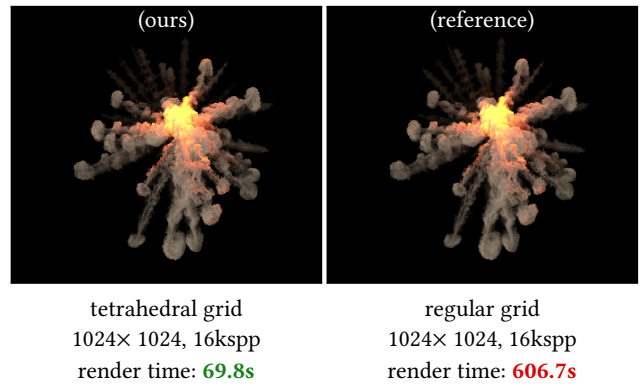

To render our volumes, we wrote a progressive path tracer in DirectX 12 that solely consists of a 
compute shader. We showcase a rendering result in Figures~\ref{fig_teaser},~\ref{fig_explosion} and in our supplemental videos. 
The cloud rendered in Figure~\ref{fig_teaser} is an OpenVDB volume and part of the Walt 
Disney Animation Studios \textit{cloud} dataset (CC-BY-SA 3.0). We render the grid reference 
using an 1024$\times$1024$\times$1024 regular grid. In contrast, our adaptive grid carries 
5,393,482 tetrahedrons (so roughly $\times$200 less cells) for this particular shot. Thanks to our 
grid's adaptive nature, we render this shot 30 times
faster than a regular grid on all consumer grade GPUs we tested. The rendering cost scales linearly with the 
number of samples, and at 32 samples per pixel, we are able to render the asset in less than 
50ms, which is fast enough for real-time pre-visualization. We observe similar performance benefits 
on our explosion dataset shown in Figure~\ref{fig_explosion}.

\bibliographystyle{ACM-Reference-Format}
\bibliography{leb3d} 

\end{document}